\documentclass[12pt]{article}
\usepackage{epsf}
\title{\bf Field distributions in heavy  mesons and baryons}
\author{
D.S.Kuzmenko\thanks{e-mail: kuzmenko@heron.itep.ru}, Yu.A.Simonov
\thanks{e-mail: simonov@heron.itep.ru}}
\date{\it Institute of Theoretical and Experimental Physics,\\
117218, Moscow, Russia}
\newcommand{\be} {\begin{equation}}
\newcommand{\ee} {\end{equation}}
\newcommand{\bdm} {\begin{displaymath}}
\newcommand{\edm} {\end{displaymath}}
\newcommand{\bc} {\begin{center}}
\newcommand{\ec} {\end{center}}
\newcommand{\beqa} {\begin{eqnarray}}
\newcommand{\eeqa} {\end{eqnarray}}

\newcommand{\ver}{\mbox{\boldmath${\rm r}$}}
\newcommand{\veE}{\mbox{\boldmath${\rm E}$}}

\begin{document}
\maketitle

\begin{abstract}
Field distributions generated by static $Q\bar Q$ and $QQQ$ sources
are calculated analytically in the framework of the Field Correlator
Method (FCM) using Gaussian (bilocal) correlator. In both cases the
string consists mostly of  longitudinal color electric field, while
transverse electric field contributes locally less then 3\%, in
agreement with earlier lattice studies. In the $QQQ$ case the profile
of the $Y$ shape
was calculated for the first time and found to have a complicated
structure with a deep well at the string junction position. Possible
consequences of this form  for the baryon structure are discussed.
\end{abstract}

\section{Introduction}
 Field distributions inside the string connecting static $Q\bar Q$
sources have been measured repeatedly on the lattice using both
connected [1-3] and disconnected \cite{4,5} probes. Similar
measurements were done later also for Abelian projected
configurations \cite{6}. Analytic calculations for the disconnected
probe made in \cite{7} in the framework of the Gaussian
approximation to the FCM
 \cite{8,9} have revealed a clear
string-like structure of the same type as was found on the lattice.

However the connected probe yields an independent and more direct
information on the field distribution in the string as compared to
 the disconnected one, namely the expectation
values of the fields themselves rather then that of field squares.

A detailed comparison of lattice data \cite{1}
 with analytic predictions of FCM was done in \cite{2}, demonstrating
 a remarkable agreement in all distributions. In particular, the
 measured decrease of longitudinal electric field $E_{\parallel}$
 with distance from  the string axis ("the string profile") agrees
 remarkably well with the contribution of the lowest, (bilocal)
 correlator \cite{2}.  It should be noted that the input of FCM is
 the field correlator as a function of distance, which is taken from the
 lattice measurements  \cite{10}, yielding an
 exponential form of both  scalar formfactors $D$ and $D_1$ \cite{8}
 with the slope $T_g\approx 0.2 $fm. The dominance of the bilocal
 correlator (sometimes called the Gaussian Stochastic Model (GSM) of
 the QCD vacuum) was verified recently on the lattice by the precision
 measurement of  Wilson loops (static potentials) in different SU(3)
 representations \cite{11}. Analysis of data 
 \cite{11} made in \cite{12} has demonstrated that GSM contribution 
 around 99\% to the static  $Q\bar Q$ potential is consistent with the data.
  
 These results give an additional stimulus to the analytic
 calculation of field distributions using lowest bilocal correlator.
 Our purpose in this letter is to calculate $3d$ field distributions
 to demonstrate the string formation, and in  particular to study
 transverse electric fields and nonconfining piece $(D_1)$ of the
 bilocal correlator which was not done in \cite{2}. In addition we
 calculate for the first time the field distributions in the static
 $QQQ$ system as a function of distance between quarks forming an
 equilateral triangle.
 It will be shown that for the lowest energy configuration
  (the so-called $Y$-shaped, or
 string-junction configuration) the  field distribution has a rather
 peculiar "well" form and vanishes at the place of the string-junction.
  The analysis of  equations reveals the reason of such shape, and
 yields an estimate of the well size.

 The plan of
 the paper is as follows. In section 2  necessary formulas are
 derived in the framework of FCM for the connected-probe field
 distributions. The $Q \bar Q$ case is treated in detail
 and $3d$ distributions are given. In section 3 the $QQQ$ case is
 considered and the relief of the $Y$-shaped string with the well is given. 
 In section 4 the $3Q$ potential is 
 derived and compared with lattice measurements. In conclusion 
 possible physical consequences and  relation to the real case of
 heavy and light hadrons are discussed.

\section{ $Q \bar Q$ case} 
 We follow in this section notations and methods used previously in
\cite{2}. A gauge invariant connected probe $\rho_{\mu\nu} (x)$ is made with the
infinitesimal  plaquette $P_{\mu\nu}$ at the point $x$, connected by
parallel transporters (Schwinger lines) $L$, $L^+$ to the contour $C$
of the Wilson loop $W$ as shown in Fig.1.

\begin{figure}[t] 
\epsfysize=6cm 
\centering
\epsfbox{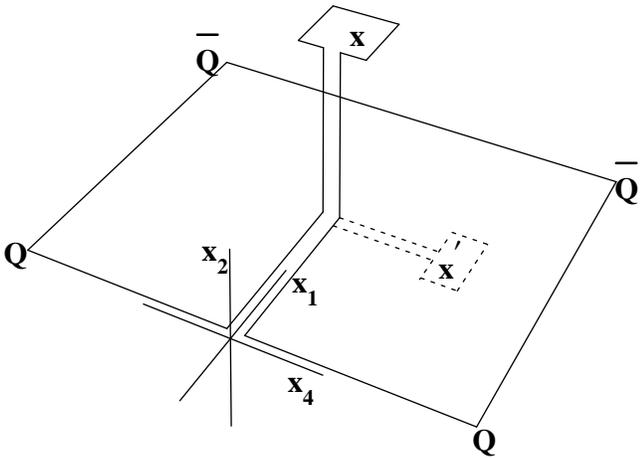} 
\caption{A connected probe in the $Q\bar Q$ case.}
\end{figure}

  \be \rho_{\mu\nu} (x) =\frac{\langle \mathrm{tr} (
WLP_{\mu\nu}(x) L^+)\rangle }{\langle \mathrm{tr} W\rangle} -1 \equiv \langle
P_{\mu\nu} (x) \rangle _{Q\bar Q}.
\label{1}
 \ee
  Schwinger lines contribution to probe depends on their length and shape
  and is minimal when they are taken to be straight lines as shown in Fig.1.
 When the size  $a$ of the plaquette $P_{\mu\nu}$ tends to zero one has
\be
 \langle P_{\mu\nu} (x) \rangle_{Q\bar Q} \to a^2 \langle
F_{\mu\nu} (x) \rangle_{Q\bar Q}.
  \label{2}
   \ee

To calculate (\ref{1}) one can use the cluster expansion theorem
\cite{8} applied to the complex Wilson loop $\bar C$ consisting of
the original contour $C$ and the plaquette $P_{\mu\nu}$ attached with
the lines $L, L^+$. In the Gaussian (bilocal) approximation one can proceed as
in \cite{2} to obtain the final result for the contribution of the lowest
correlator, $ \langle FF\rangle$, \be
\rho_{\mu\nu} (x) = a^2 \int_{S(C)}d\sigma_{\rho\lambda} (x')
D_{\rho\lambda,\mu\nu} (x',x)+O(a^4).
\label{3}
\ee
Here $S(C)$ is the minimal area surface inside the contour $C$, and
$D$ is
\be
D_{\rho\lambda,\mu\nu} (x',x) =\frac{g^2}{N_c} \mathrm{tr} \langle
F_{\rho\lambda}(x') \Phi (x', x) F_{\mu\nu} (x) \Phi(x,x')\rangle,
\label{4}
\ee
 while  $\Phi(x,y)$ is the parallel transporter, connecting points $x,y$.
 Integration goes over infinitesimal plaquettes 
 $\Delta \sigma_{\rho\lambda}(x')$ on the surface $S$ as depicted in Fig.1 by
 dashed line. 
 For the gauge-invariant function $D_{\rho\lambda, \mu\nu}$ there
 exists an expression in terms of two scalar functions $D$ and $D_1$
 (see second ref. in \cite{8}),
 $$
D_{\rho\lambda,\mu\nu} (x,y ) = (\delta_{\rho\mu}
\delta_{\lambda\nu}-\delta_{\rho\nu}\delta_{\mu\lambda})
D (h)+
$$
\be
\frac12 [\frac{\partial}{\partial h_\rho}
h_\mu\delta_{\lambda\nu} -
\frac{\partial}{\partial h_\lambda}
h_\mu\delta_{\rho\nu} + \rho \lambda\leftrightarrow \mu\nu] D_1(h),
\label{5}
\ee
where $h\equiv x-y$. The functions $D, D_1$ have been measured on the
lattice \cite{10} and were found to have an exponential form beyond
$x=0.1$fm.

Here we use the exponential ansatz in the whole region of
$x$, as it was done previously in \cite{2}, with parameters
as in \cite{10},\cite{2}
\be
D(x) = D(0) \exp (-\mu|x|),
~~D_1(x) = D_1(0) \exp (-\mu|x|),
\label{6}
\ee
$$
D_1(0) \approx \frac13 D(0),~~ \mu\approx 1 \mathrm{GeV} .$$

In the same Gaussian approximation, i.e. neglecting all higher
 correlators  (which is consistent with lattice measurements of static
 potentials with accuracy of better than one percent  \cite{11,12}) $D(0)$
  can be expressed through
 the string tension \be \sigma =\frac12 \int D(x) d^2 x= \pi
D(0)/\mu^2.  \label{7} \ee This connection will be important in what
 follows for the correct normalization of the field distribution in
 $Q\bar Q$ and $QQQ$.

 To proceed with the integration in (3), choose the coordinate axes as shown
 in Fig.1. The contour $C$ of the Wilson loop is a rectangular 
 of size $R\times T$ where $R$ is quark separation and $T$ is temporal 
 extension of the Wilson loop. Plaquettes $\Delta \sigma_{\rho\lambda}(x')$,
 $x'=(x_1',x_2',x_3',x_4')$, have coordinates $0\leq x_1' \leq R,
  x_2' =x_3' =0, -T/2\leq x_4' \leq T/2$. The probe is placed at
 $x=(x_1,x_2,x_3,x_4)$. $x_1$ is the coordinate of the probe along string axis,
 $x_2$ is the distance from the probe to the string axis, $x_3=x_4=0$. The only
 nonzero components of color field are electric ones in $x_1$ and $x_2$ 
 directions (see \cite{2} for symmetry arguments and discussion).
 The resulting equations are directly obtained from (\ref{2}-\ref{5})
 ($\rho_{i4}(x_1,x_2)= a^2 \langle E_i(x_1,x_2) \rangle_{Q\bar Q}
 \equiv a^2 E_i(x_1,x_2)$):
 \be
 E_1 (x_1, x_2) = \int^R_0 dx'_1
 \int^{\frac{T}{2}}_{-\frac{T}{2}} dx'_4 [D(0) +D_1(0)
 -\mu\frac{h^2_4+h^2_1}{2h} D_1 (0)]e^{-\mu h},
 \label{8}
 \ee
 \be
 E_2 (x_1, x_2) = - \mu x_2 \int^R_0 dx'_1
 \int^{\frac{T}{2}}_{-\frac{T}{2}} dx'_4
 \frac{(x_1-x'_1)}{2h}
  D_1 (0) e^{-\mu h}.
 \label{9}
 \ee

 Here we have defined
 \be
 h_4=-x'_4,~~ h_1 =x_1-x'_1; ~~ h^2= h^2_4+ h^2_1+x^2_2.
 \label{10}
 \ee
 The field distributions $E^2_1(x_1,x_2)$ are shown in
 Fig.2 (a-c) for different quark separations, $ R= T_g, 5T_g$ and $15T_g$.
 The string profile, i.e. $E^2_1(x_2)$ distribution in the middle of the string, for these quark 
  separations is shown in Fig.2 (d). One can see a clear string-like  formation with the  width of $2.2T_g$.
 The value of $E_1$ in the middle of the string does not depend on $R$ for $R > 5T_g$.
 The saturated value is $E^{sat}=1.8$GeV/fm.
 
The field distribution $E_2(x_1,x_2)$ is shown in Fig.3\footnote{qualitative
figures 3,4,6 at http://heron.itep.ru/$\sim$kuzmenko/figs.uu are 
available} 
for $R=15T_g$. Note that $|E_2|$ is much smaller than $|E_1|, 
 \max |E_2(x_1,x_2)| \approx 0.03 \max |E_1(x_1,x_2)|$.

The distribution $E_1(x_1,x_2)$ generated by
$D_1(x)$ is shown separately in Fig.4. One can see that $D_1$ does not produce a string, which is
in agreement with the fact, that $D_1$ does not contribute to the
string tension (\ref{7}).

\section{$QQQ$ case}
 In case of  the $QQQ$ system the contour $C$ consists of three
contours connected by the string-junction trajectory, as depicted in
Fig.5.  The surface $S$ consists of three surfaces $\Sigma_A,
\Sigma_B$ and $\Sigma_C$, and the probing
plaquette $P_{\mu\nu}$ is attached by Schwinger lines $L, L^+$ to one
of the static quark trajectories.

\setcounter{figure}{4}
\begin{figure}[t] 
\epsfysize=6cm 
\centering
\epsfbox{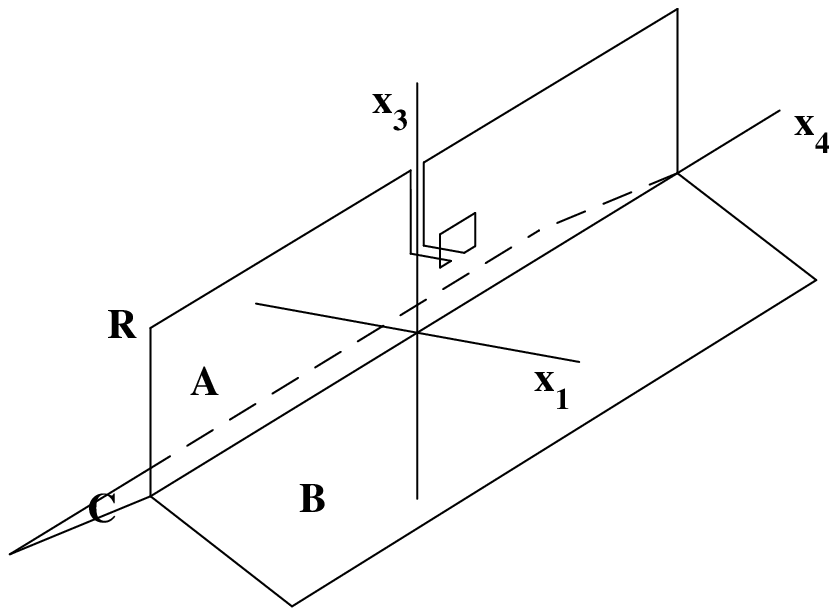} 
\caption{A connected probe in the $QQQ$ case.}
\end{figure}

Choosing the axes as shown in Fig.5, $x_2=0$, one has two main contributing
correlators $D_{14, 14}$ and $D_{34,34}$ and a subleading
nondiagonal one $D_{14,34}= D_{34,14}$.  As a result one obtains the
following field distributions (choosing $x_4=0$ for the plaquette $P_{\mu\nu}$
as in the $Q \bar Q$ case):
\be
E^{(3Q)}_1 (x_1,x_3) = \int
(d\sigma_A\Lambda^{(A)}_{14} +d \sigma_B \Lambda_{14}^{(B)} +
d\sigma_C\Lambda^{(C)}_{14}),
\label{11}
\ee
\be
E^{(3Q)}_3 (x_1,x_3) = \int
(d\sigma_A\Lambda^{(A)}_{34} +d \sigma_B \Lambda_{34}^{(B)} +
d\sigma_C\Lambda^{(C)}_{34}),
\label{12}
\ee
where we have defined
\be
d\sigma_A\Lambda^{(A)}_{14} = dx'_4dx'_3 D_{34,14}(x-x'),
\label{13}
\ee
\be
d\sigma_B\Lambda^{(B)}_{14} =\frac{\sqrt{3}}{2}  dx'_4dl'
D_{14,14}-\frac12 dx'_4 dl' D_{34,14},
 \label{14} \ee
\be
d\sigma_C\Lambda^{(C)}_{14} =-\frac{\sqrt{3}}{2}  dx'_4dl'
D_{14,14}-\frac12 dx'_4 dl' D_{34,14},
 \label{15}
  \ee
\be
d\sigma_A\Lambda^{(A)}_{34} =  dx'_4dx'_3
D_{34,34}(x-x'),
 \label{16}
  \ee
\be
d\sigma_B\Lambda^{(B)}_{34} =\frac{\sqrt{3}}{2}  dx'_4dl'
D_{14,34}-\frac12 dx'_4 dl' D_{34,34},
 \label{17}
  \ee
\be
d\sigma_C\Lambda^{(C)}_{34} =-\frac{\sqrt{3}}{2}  dx'_4dl'
D_{14,34}-\frac12 dx'_4 dl' D_{34,34}.
 \label{18}
  \ee

  Here $dl'$ denotes integration from $l'=0$ to $l'=R$, $R$ is quark 
  separation from the string junction; $dx_4'$ denotes integration from 
  $x_4'=-T/2$ to $x_4'=T/2$.

  Using  (\ref{5}) the correlators entering
  (\ref{13}-\ref{18}) can be  expressed through $D, D_1$ as
  \be
  D_{14, 14 }(z) = D(z) +D_1 (z)+(z^2_1+z^2_4)\frac{dD_1}{dz^2},
  \label{19}
  \ee
  \be
  D_{34, 34 }(z) = D(z) +D_1 (z)+(z^2_3+z^2_4)\frac{dD_1}{dz^2},
  \label{20}
  \ee
  \be
  D_{14, 34 }(z) = D_{34,14}(z)= z_1z_3\frac{dD_1}{dz^2}.
   \label{21}
  \ee

  From the symmetry of Fig.5 it is clear, that the resulting field
  distribution is symmetric under  rotation in the $(x_3, x_1)$ plane
  over angle $\frac{2\pi n}{3},~~ n=0,1,2,...$ , and indeed from
  (\ref{11}-{21}) one can conclude that color electric field has this
  symmetry.

  We have calculated the $\veE^2(\ver^2)=(E^{(3Q)}_1)^2+(E^{(3Q)}_3)^2$ 
  distribution, where $\ver =(x_1,x_3),$ displayed on Fig.6 (a-c) for 
  $R=T_g,5T_g$ and $15T_g$ respectively.  One can easily see the symmetry 
  discussed above and in addition a remarkable feature - a well in 
  $\veE^2(\ver^2)$ near $\ver=0$. The well profile, i.e. distribution of 
  $\veE^2$ along the $x_3$ axis, is depicted in Fig.6 (d) for given quark
   separations. As one can see, the well shape does not depend on the quark 
   separation. At separation $15T_g$ the string acquires its saturation
   value $E^{sat}=1.8$GeV/fm. The radius of the well,  i.e. distance from
  the string  junction to the position at which $\veE^2$ is half of its 
  saturation value, is  $1.75T_g$.
  
  The physical origin of this well is understandable: the fields
  $\veE$ are directed in three sheets $\Sigma_A, \Sigma_B, \Sigma_C$
  and are perpendicular to the string-junction line; hence  there
  is no preferred direction for the field $\veE$ at $\ver=0$. To get
  more quantitative insight into the problem, let us consider the
  line $x_1=0$ in the plane $(x_1, x_3)$ and compute
  $E_1^{(3Q)}$ and $E^{(3Q)}_3$ as functions of
  $x_3$. From (\ref{11}), (\ref{12}) one immediately obtains
  \be
  E^{(3Q)}_1 (x_1=0, x_3) =0,
  \label{22}
  \ee
  $$
  E^{(3Q)}_3 (x_1=0, x_3)
  =\int^{\frac{T}{2}}_{-\frac{T}{2}} dx'_4 \int^R_0 dl'
  \{\tilde
  D(\sqrt{(x'_4)^2+ (x_3-l')^2})
  $$
  \be
    -\tilde
  D(\sqrt{(x'_4)^2+\frac34l^{'2}+ (x_3+\frac12 l')^2}) \},
   \label{23} \ee
  where $\tilde D\equiv D_{34,34}(h)$.
  From (\ref{23}) one can deduce that $E_3$ vanishes linearly
  at $x_3=0$ and grows fast for  positive $x_3$ attaining a constant
  limit for  $x_3\sim \frac{1}{\mu}$ when $R\mu\gg 1$.
  Thus $\veE = 0$ at the center of the well and the radius of the
  well is of the order of $r\sim \frac{1}{\mu}=T_g$. 

 Using the connected probe we have built the square of expectation value 
 of baryon field from the expectation values of fields of 
 Wilson loop three sheets.
 Let us mention that by the disconnected  probe we could measure the
 expectation value of squared fields and would know nothing about fields
 themselves since in general $<E^2>\neq <E>^2$.

  \section{Quark interaction potential in $QQQ$ case}
  We now go over to another topic, closely connected with the
  previous one, namely the static potential for the $QQQ$ system.
  Choosing the same  $Y$-shaped equilateral configuration one has
  from the cluster expansion of the Wilson loop $W^{(3Q)}$,
  $$
  V^{(3Q)}(R) =- \lim_{T\to \infty} \frac{1}{T}\ln \langle
  W^{(3Q)}(R,T)\rangle=
  $$
  \be
  =\sum_{a,b=A,B,C}\frac12 \int\int d\sigma_a(x) d\sigma_b(y)
  D_{a,b}(x,y) =
  \frac12 \sum_{a,b}V_{a,b}(R),
  \label{24}
  \ee
  where, e.g.
  \be
  V_{A,A}(R) =\frac{1}{T}
  \int^{\frac{T}{2}}_{-\frac{T}{2} } dt
  \int^{\frac{T}{2}}_{-\frac{T}{2} } dt'
  \int^R_0 dz
  \int^R_0 dz'
  [D(h)+
  D_1+h^2\frac{dD_1}{dh^2}]
  \label{25}
  \ee
  and $h^2=(z-z')^2+(t-t')^2$.

  Using (\ref{13})-(\ref{18}) one can easily find the nondiagonal terms, e.g.
  \be
  V_{A,B}(R) =\frac{1}{T}
  \int^{\frac{T}{2}}_{-\frac{T}{2} } dt \int^{\frac{T}{2}}_{-\frac{T}{2}}dt'
  \int^R_0 dl
  \int^R_0 dl'
  [\frac{\sqrt{3}}{2}D_{14,34}(\tilde h)-\frac12
  D_{34,34}(\tilde h)],
   \label{26} \ee
  where $\tilde h^2=(t-t')^2+\tilde h^2_1+\tilde h^2_3,~~ \tilde
  h_1=-\frac{\sqrt{3}}{2} l', ~~\tilde h_3=l+\frac12 l';$
  and
  \be
  D_{14,34}(\tilde h) =\tilde h_1\tilde h_3\frac{dD_1(\tilde
  h^2)}{d\tilde h^2}.
  \label{27}
  \ee
  For $V_{A,C}(R)$ both (\ref{26}) and (\ref{27}) are valid, but the
   sign of $\tilde h_1$ is opposite to  that of (\ref{27}).

    From symmetry of the problem it is clear that the total potential
  is expressed through $V_{A,A}, V_{A,B}, V_{A,C}$ as
  \be
  V^{(3Q)}(R)=\frac{3}{2} (V_{A,A}+V_{A,B}+ V_{A,C}).
  \label{28}
  \ee
  We plot this potential in Fig.7 together
  with the total potential including the perturbative
  one-gluon-exchange parts,
  \be
  V_{tot}^{(3Q)}(R)=V^{(3Q)}(R) -\frac{2\alpha_s}{\sqrt{3}R},
  \ee
  where $\alpha_s=3/4e$. $e=0.295$ from fit of the $Q\bar Q$ lattice 
  potential by the Coulomb plus linear (Cornell) potential (see \cite{13} 
  and references therein). 
  One can notice in Fig.7 that large-distance asymptotic of the
  $V^{(3Q)} (R)$ is equal to $3\sigma R$, as it should be. 
  At $R=3.5T_g$ the slope of $V^{(3Q)}$ is $2.6\sigma$. This is 
  consistent with recent lattice  measurements of $V^{(3Q)}_{tot} (R)$
  (\cite{13}) fitted in the measured region $0.055$fm$\leq R\leq 0.71$fm  
  by the Cornell potential with the same slope of linear part.

  \setcounter{figure}{6}
  \begin{figure}[t] 
  \epsfysize=6cm 
  \centering
  \epsfbox{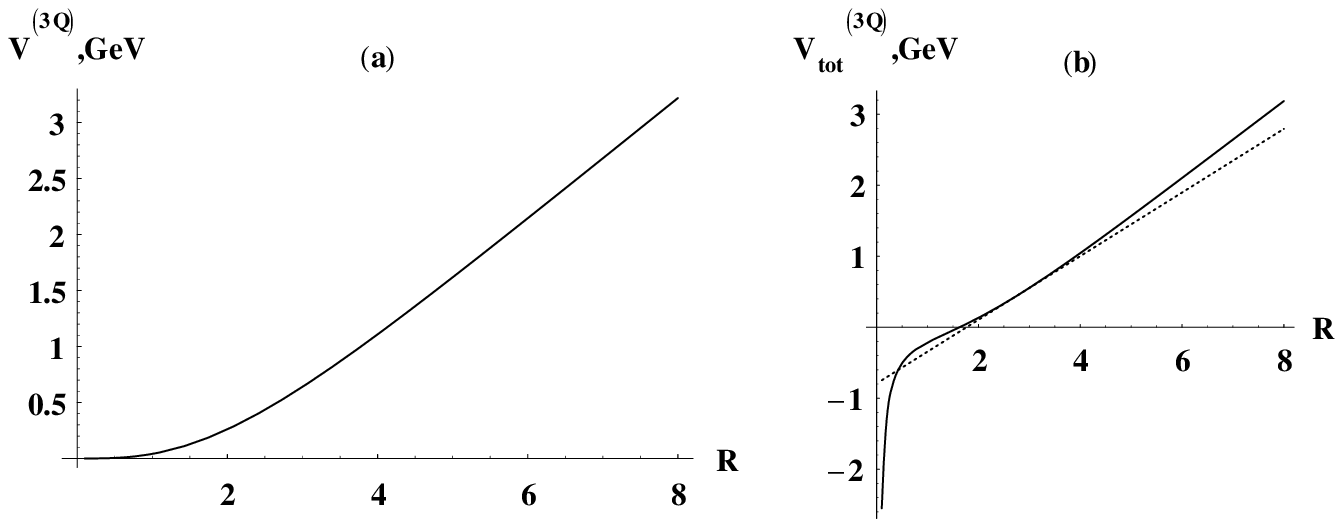} 
  \caption{(a).  $3Q$ potential for the string junction configuration
  of Fig.5 as a function of distance from quark to string junction (in units of $T_g$).
  (b). The same as in (a) with perturbative one-gluon-exchange potential added.}
  \end{figure}

   \section{Conclusions}  
   We have calculated connected-probe field distributions for
   $Q\bar Q$ and $QQQ$ cases using the lowest (Gaussian) field
   correlator. Basing on the results of [11,12],   one expects that
   higher correlators would not significantly change our picture. 

   The connected-probe analysis is essential for establishing
   direction of fields in the string and distinguishing
   color-electric and color-magnetic contributions, since due to
   properties of the Gaussian correlator (\ref{5}) the orientation of the
   probing plaquette
   $P_{\mu\nu}$ allows to fix the vector of the field in the string.

   Our results for the $Q\bar Q$ case are in agreement with earlier
   calculations in [1-3]. The bulk of the string fields is also
   in accordance with disconnected-probe analyses \cite{4,5}.

   The QQQ results are new and striking. A deep well in the electric
   field distribution appears at the string-junction position.
   Exactly at this point electric field vanishes  because of rigorous 
   symmetry arguments. The well has a radius of $1.75T_g$ and implies strong
   suppression of fields in the middle of the heavy baryon. Since the
   Wilson  loop for the QQQ configuration can be used to generate
   interaction applicable for light quarks \cite{14}, physical
   consequences of this suppression can be in principle observable both for
   light and heavy hadrons. One consequence can be seen in Fig.7,
   where the nonperturbative  part of the potential grows very slowly
   at small $R$, so that the asymptotic slope is obtained only at
   very large distances. Therefore an effective  slope for
   ground-state hadrons can be some 10-20\% smaller, the fact which is in
agreement with relativistic quark    model of baryons \cite{15} and with
recent lattice calculations    of static QQQ potential \cite{13}.

   One should note that the vanishing of fields at the string
   junction holds for \underline{directed field}\\ 
   \underline{distribution}, measured
   in the connected-probe analysis, and may not be true for field
   fluctuations, measured in the disconnected probe. This topic and
   other possible consequences of the field suppression in the middle of the
   baryon call for further investigations.

   Financial support of the RFFI grants 00-02-17836 and 00-15-96786 is gratefully
   acknowledged.

   The authors are grateful to N.O.Agasyan, A.B.Kaidalov, Yu.S.Kalashnikova,
   V.I.Shev-\\ chenko for useful remarks. D.K. is grateful to D.V.Chekin for
   a discussion of software details.

\enlargethispage{\baselineskip}

\setcounter{figure}{1}
\begin{figure}[h] 
\epsfysize=13cm 
\centering
\epsfbox{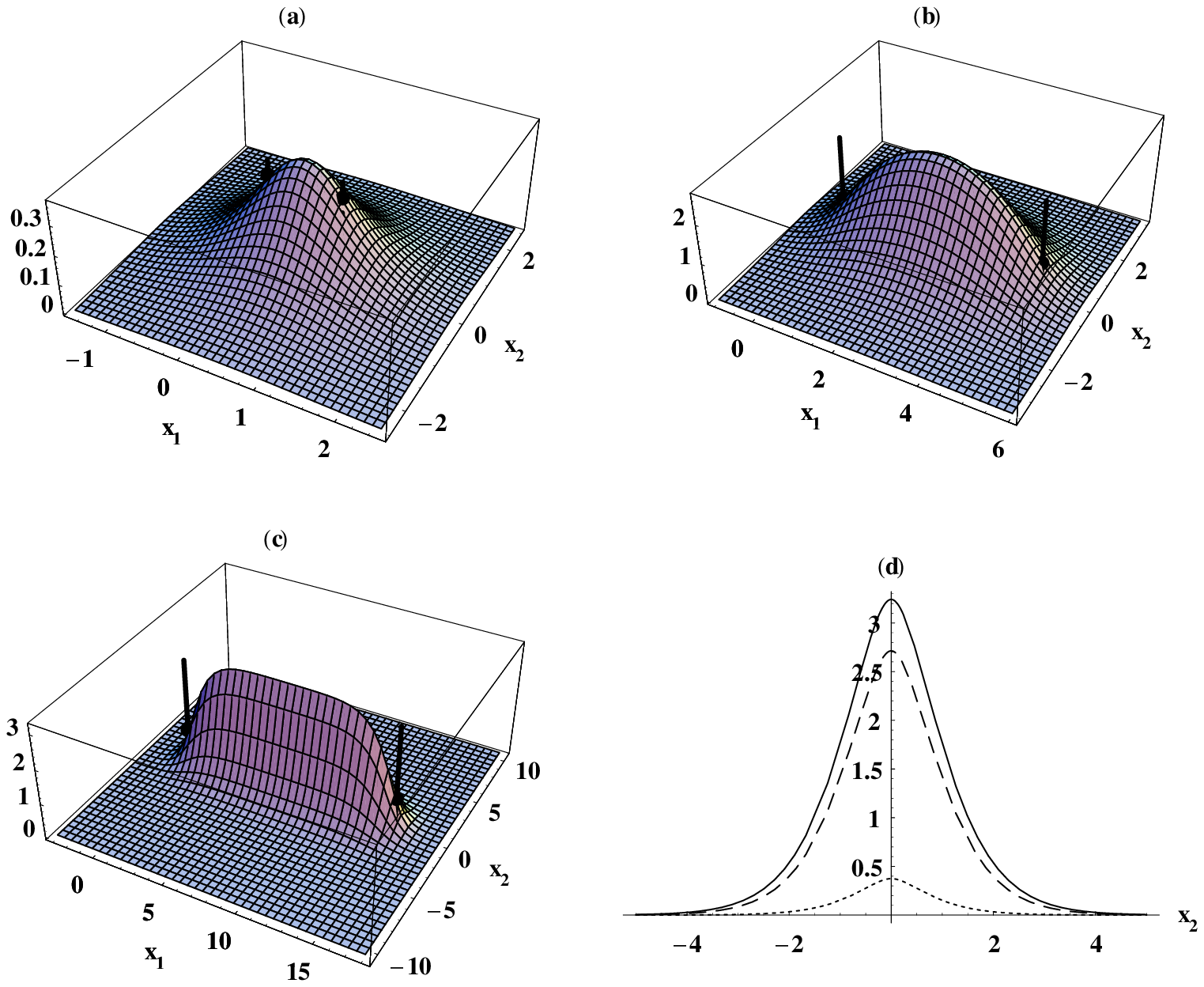} 
\caption{Distribution of the field $E_1^2$ measured in GeV$^2$/fm$^2$ around 
static $Q\bar Q$ sources
as function of $x_1$ (distance along the string) and $x_2$ (distance from the
string axis). Both distances are measured in units of $T_g$. Positions of $Q$ and
$\bar Q$ are marked with points with vertical lines. The cases (a),(b),(c) 
refer to the interquark distances $R_{Q\bar Q}=1,5$ and $15T_g$. In Fig.2(d)
the string profile (i.e. $E_1^2(x_2)$ distribution at the middle of the string)
is shown,
dotted, dashed and solid lines for the three interquark distances of Figs.
2(a), (b) and (c) respectively. Note the saturation of the string height and
width at around $R \approx 5T_g$.}
\end{figure}

\end{document}